\documentclass[12pt]{iopart}
\usepackage{graphicx}
\usepackage[dvipsnames]{xcolor}

\begin{document}

\title{Nonresonant capture cross section measurement of the $^{29}$Si(p,$\gamma$)$^{30}$P reaction}

\author{Zs M\'atyus$^{1,2}$, L Csedreki$^1$, Zs F\"ul\"op$^1$, Z Hal\'asz$^1$, G G Kiss$^1$, T~Sz\"ucs$^1$, \'A~T\'oth$^{1,2}$, Gy Gy\"urky$^1$\footnote{corresponding author, gyurky@atomki.hu}}%
\address{$^1$HUN-REN Institute for Nuclear Research (ATOMKI), 4001 Debrecen, Hungary}
\address{$^2$University of Debrecen, Doctoral School of Physics, Egyetem t\'er 1., 4032 Debrecen, Hungary}

\begin{abstract}

The isotopic ratios measured in meteoritic presolar grains are a crucial tool for tracing the nucleosynthetic origin of isotopes. In the case of silicon isotopes, two important indicators to establish the origin of presolar grains are the ratios $^{29}$Si/$^{28}$Si and $^{30}$Si/$^{28}$Si. To constrain theoretical predictions, the rates of key nuclear reactions influencing the abundances of $^{29}$Si and $^{30}$Si must be well known. One such reaction is $^{29}$Si(p,$\gamma$)$^{30}$P which plays a role in classical nova explosions. The aim of the present work is to determine the nonresonant cross section of the $^{29}$Si(p,$\gamma$)$^{30}$P reaction, which has not been previously measured. The activation method was employed to measure the total cross section at four proton energies between $E_p$\,=\,1000 and 1430 keV.  The measured cross sections were found to be significantly (a factor of 4.3\,$\pm$\,0.6) higher than those predicted by theoretical direct capture calculations, thereby impacting the reaction rates at low astrophysical temperatures, below about 30\,MK. This higher nonresonant cross section - now based on experimental data - can be used in forthcoming nucleosynthesis calculations of classical novae. As a secondary result, the $^{16}$O(p,$\gamma$)$^{17}$F cross section was also obtained and found to be in good agreement with existing literature data. 

\end{abstract}

\submitto{\JPG}
%\maketitle

\section{\label{sec:intro} Introduction}
Thermonuclear runaway (TNR) events such as classical nova explosions are prolific dust producers \cite{Jose2007,Jose2016}. The produced dust may become enclosed in meteoritic grains within primitive meteorites. The isotopes created by TNRs are preserved in these grains, along with their relative abundances. By measuring these isotopic ratios, valuable insights can be gained into the nucleosynthesis processes occurring in TNR scenarios, provided that the meteoritic grain's origin from a TNR event is unambiguously identified. Up to now, the nova paternity of these grains is not clearly proven, supernovae are often considered as an alternative source \cite{Downen2022b}. However, 18 grains with a likely nova origin have been recently identified in the current inventory of presolar grains \cite{Iliadis2018}.

One reason behind this uncertainty is the insufficient knowledge of the thermonuclear reaction rates of the key reactions in the production and destruction of some key isotopes. In the case of silicon isotopes, -- according to the calculations of e.g. Downen et al. \cite[Section 2.]{Downen2022b} -- the $^{29}$Si/$^{28}$Si and the $^{30}$Si/$^{28}$Si isotopic ratios are mainly determined by the $^{29}$Si(p,$\gamma$)$^{30}$P and $^{30}$P(p,$\gamma$)$^{31}$S reactions. Nova models provide values for these ratios, expressed as $\delta^{29,30}$Si/$^{28}$Si = [($^{29,30}$Si/$^{28}$Si)/($^{29,30}$Si/$^{28}$Si)$_\odot$ - 1] $\times$ 1000, or deviations from solar in permil,  in the range  $\delta^{29}$Si/$^{28}$Si = -900 to 10  $\delta^{30}$Si/$^{28}$Si = -1000 to 47000 \cite{Jose2004}. A sensitivity study done by Iliadis et al. \cite{Iliadis2002} indicated that the uncertainties of $^{29}$Si(p,$\gamma$)$^{30}$P and $^{30}$P(p,$\gamma$)$^{31}$S reaction rates strongly contribute to the uncertainties of the silicon isotopic ratio predictions in classical nova models. 

In novae, synthesis of intermediate-mass elements in the range from Si to Ca, is dominated by the very influential reaction $^{30}$P(p,$\gamma$)$^{31}$S \cite{Jose2001}. Accordingly, it is important to understand how much $^{30}$P can be synthesized during a nova outburst, reinforcing the interest in improving the knowledge of the $^{29}$Si(p,$\gamma$)$^{30}$P rate.

At nova temperatures (T = 0.1 - 0.4 GK), the $^{29}$Si(p,$\gamma$)$^{30}$P reaction rate is dominated by some low-energy resonances. Recently, the strengths of some of these resonances were measured by Downen et al. \cite{Downen2022}. The strengths were determined relative to the $E_{\rm p}$\,=\,416.9\,keV resonance, which had ambiguous strength in the literature. In order to remedy this ambiguity, this strength was measured precisely by our group using the activation method \cite{Matyus2024}. Owing to this new result, the strengths of the astrophysically important low-energy resonances have now more reliable values.

The direct capture (DC) component of the $^{29}$Si(p,$\gamma$)$^{30}$P  cross section is, on the other hand, not known at all experimentally, as no measurement was carried out so far for this quantity. Although the direct capture is dominant in the reaction rate only at the lowest temperatures below about 30\,MK \cite{Iliadis2001}, its knowledge is needed for the full understanding of this reaction and for the reliable calculation of its reaction rate at low temperatures. 

The missing experimental DC cross section is not only true for the $^{29}$Si(p,$\gamma$)$^{30}$P reaction, but it is a more general problem in this mass range. The main reason is that at low energies the cross sections are typically too low for the currently available experimental techniques, while at higher energies resonances often dominate the cross section. Therefore, for most reactions, DC cross section is estimated by scaling single-particle calculations
with spectroscopic factors (or ANCs) from transfer experiments \cite{Longland2010}. The lack of experimental data hinders the test of these theoretical calculations, making the DC cross section a source of reaction rate uncertainty. 
Therefore, the aim of the present work is to measure the DC reaction cross section of $^{29}$Si(p,$\gamma$)$^{30}$P where, with the carefully chosen proton energies, nonresonant cross section could be measured for the first time.

\section{\label{sec:experimental} Experimental procedure}

Since in our recent experiment the activation method proved to be suitable for studying the $^{29}$Si(p,$\gamma$)$^{30}$P reaction \cite{Matyus2024}, this technique was used in the present work to measure the nonresonant capture cross section. The term ``nonresonant capture'' is used as the measured cross section may include not only the DC cross section, but possible contribution of e.g. low-energy tails of broad resonances, as it will be discussed in sec \ref{sec:discussion}. The applied activation method is based on the decay counting of the $^{30}$P reaction product, and provides thus the total reaction cross section, independently from the de-excitation scheme of $^{30}$P after the proton capture of $^{29}$Si. The total cross section is the astrophysically relevant quantity as it is needed for the reaction rate determination. The following subsections describe this experimental procedure in detail. 

\subsection{\label{sec:target}Target properties}
The targets were made by thermal vacuum evaporation. Since the natural abundance of $^{29}$Si is only about 4\%, highly enriched $^{29}$Si (99.34\,\% isotopic enrichment, purchased from ISOFLEX USA) was used. The Si layer was evaporated onto 0.5 mm thick high-purity (99.9\,\%) Ta sheets. Several targets were prepared, some of them used for the initial tests. The results presented here were obtained using three targets, referred to as \#4, \#6 and \#9 in the following.

Although the starting material for the evaporation is pure elemental Si, oxidation takes place during or after the evaporation. Therefore, in order to determine the areal density of the $^{29}$Si target atoms - needed for the cross section calculation - the thickness of the target layers as well as the Si:O ratios need to be measured. 

The Si:O ratio was determined by measuring the yield of the $E_{\rm p}$\,=\,416.9\,keV resonance. The strength of this resonance is known from our previous study \cite{Matyus2024}, thus the effective stopping power of the target can be calculated from the measured yield which then provides the Si:O ratio. This ratio was found to be in the range of 0.75\,--\,1 for the three targets used. The relative uncertainty of these ratios is typically 12\,\%.% (see also section \ref{sec:analysis} for the details of uncertainty calculation).

\begin{figure}[ht]
   \includegraphics[clip,width=1\linewidth]{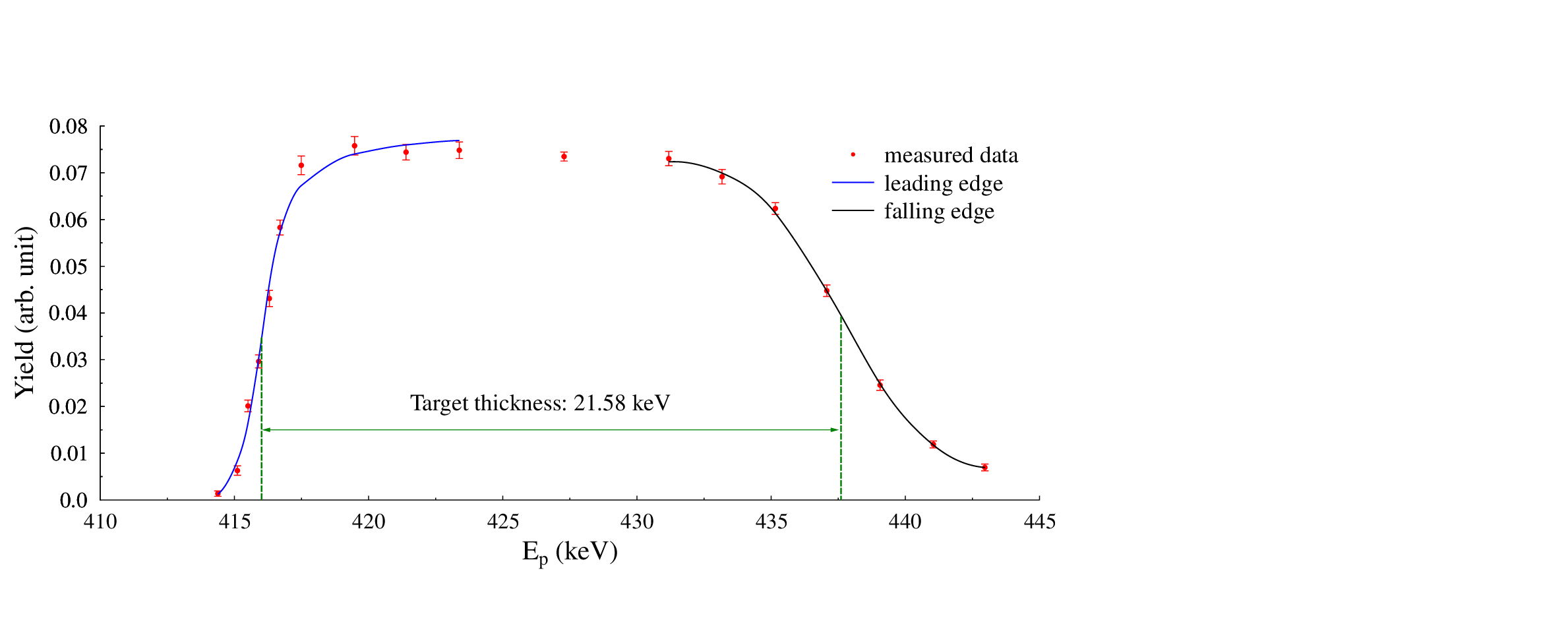}
    \caption{Resonance profile of target \#9 obtained with the $E_{\rm p}$\,=\,416.9\,keV resonance in $^{29}$Si(p,$\gamma$)$^{30}$P. From the fit of the profile as shown in the figure, the energetic thickness of the target could be obtained.}
    \label{fig:target}
\end{figure}

The thicknesses of the target layers were measured with Nuclear Resonant Reaction Analysis using again the $E_{\rm p}$\,=\,416.9\,keV resonance in $^{29}$Si(p,$\gamma$)$^{30}$P and detecting the prompt 677\,keV and 709\,keV $\gamma$-rays from the de-excitation of the first and second excited states of $^{30}$P. By scanning the proton energy around the resonance and by measuring the yield of these $\gamma$-rays, the target profile was obtained. An example of such a target profile - taken on target \#9 - is shown in Fig \ref{fig:target}. From the fit of the leading and falling edges of the resonance profile - as shown in the figure - the energetic target thickness can be determined. Knowing the Si:O ratio and thus the stopping power of the target layer, the areal density of the $^{29}$Si atoms can be calculated from the energetic target thickness. The stopping powers for O and Si are taken from the SRIM code \cite{SRIM}.

Table \ref{tab:targets} summarizes the properties of the three targets used for the cross section measurements. 

\begin{table}
\caption{\label{tab:targets} Some parameters of the used targets. The energetic thicknesses are given at 416.9\,keV proton energy. See text for further details.}
\begin{indented}
\item[]\begin{tabular}{cccc}
\br
Target no. & Si:O ratio & Energetic thickness  & $^{29}$Si areal density \\
 				&						& 		keV		&   10$^{18}$ atoms/cm$^2$ \\
\mr
\#4 & 0.81\,$\pm$\,0.13  &  24.59\,$\pm$\,0.13  &  1.027\,$\pm$\,0.078\\
\#6 & 0.77\,$\pm$\,0.13  &  22.56\,$\pm$\,0.20  &  0.919\,$\pm$\,0.070\\
\#9 & 0.99\,$\pm$\,0.14  &  21.58\,$\pm$\,0.11 &  0.987\,$\pm$\,0.074\\
\br
\end{tabular}
\end{indented}
\end{table} 

%The resonance profiles were measured both before and after the cross section measurements on each target. Fig ??? shows both profiles in the case of target ???. No difference was observed between the two profiles which prove that no target degradation occurred during the proton bombardment. 

\subsection{Irradiation} 

Since the aim of the present work was to measure the nonresonant cross section of the $^{29}$Si(p,$\gamma$)$^{30}$P reaction, the proton energies had to be chosen in such a way to avoid any known resonances within the energy range covered by the target thickness. The proton energy selection was based on the work of G Vavrina et al. \cite{Vavrina1997} who measured the yield of this reaction in an energy range between $E_{\rm p}$\,=\,1000 and 1750\,keV. No DC cross section was provided in that work, but from their list of resonances and their plotted excitation function (fig 1 in \cite{Vavrina1997}) regions devoid of resonances could be located. Based on this, four proton energies were chosen for our study: 1000, 1100, 1247 and 1430 keV. The Q-value of the $^{29}$Si(p,$\gamma$)$^{30}$P reaction is 5595 keV, thus these proton energies correspond to $^{30}$P excitation energies of 6562, 6658, 6800 and 6977 keV. According to the latest compilation \cite[page 2399]{Basunia2010}, there are no known states of $^{30}$P at or close below these energies. Thus, measuring the yield of this reaction at these energies gives a good chance to get the DC cross section (see also the discussion in section \ref{sec:discussion}). 

The proton beam was provided by the Tandetron accelerator of Atomki \cite{Biri2021}. According to the energy calibration of the accelerator, the beam energy is known with an accuracy of 0.1\,keV \cite{Rajta2018}. The target chamber was the same as in our previous work \cite{Matyus2024}. To suppress secondary electrons coming from the target or from the beam collimator, an electrode biased at -300V was placed at the entrance of the target chamber. Behind this electrode the target chamber forms a Faraday cup. The number of the protons hitting the target can thus be derived from the charge measurement, using an ORTEC 439 charge integrator. 

The beam current was typically 2.5-5$\mu$A. 
%With this beam intensity, no target degradation was observed. 
The charge integrator counts were read out in every 5 seconds which can be used to follow beam intensity variations. 

The reaction product has a short half-life of  $t_{1/2}$\,=\,2.498\,$\pm$\,0.004 min \cite{Basunia2010}, thus the so called cyclic activation was used.  The irradiation phase was 5 min long, followed by a 15 min counting phase. This 20 min cycle was repeated many times in a run, which lasted between 5 and 18 hours. Altogether 13 many-cycle runs were carried out at the four proton energies discussed above. Some relevant information about these runs are given in table \ref{tab:runs}.

\begin{table}
\caption{\label{tab:runs} Details of the many-cycle runs. The runs are listed in time order.} 
\begin{indented}
\item[]\begin{tabular}{cccc}
\br
Run & $E\rm _p$ & Target & No. of cycles  \\
& keV 				&						& 		 \\
\mr
\#1	&	1099.9	&	\#4	&	49	\\
\#2	&	999.9	&	\#4	&	57	\\
\#3	&	1247.0	&	\#4	&	51	\\
\#4	&	1427.4	&	\#4	&	60	\\
\#5	&	1427.4	&	\#6	&	49	\\
\#6	&	999.9	&	\#6	&	46	\\
\#7	&	1099.9	&	\#6	&	49	\\
\#8	&	1247.0	&	\#9	&	33	\\
\#9	&	1099.9	&	\#9	&	42	\\
\#10	&	1429.4	&	\#9	&	60	\\
\#11	&	1099.9	&	\#9	&	64	\\
\#12	&	1247.0	&	\#9	&	47	\\
\#13	&	1429.4	&	\#9	&	17	\\
\br
\end{tabular}
\end{indented}
\end{table} 

\subsection{Detection of the $^{30}$P decay} 

A HPGe detector was placed behind the chamber with its face about 20\,mm far from the target, surrounded by a lead shielding. During both the irradiation and counting phases, the detector detected the 511 keV gamma radiation following the annihilation of positrons from the decay of the $^{30}$P nuclei. Similar to the charge counting, the collected events by the detector were read out in every 5\,sec. Only the events during the beam-off counting phases were used for the analysis in order to avoid complications caused by beam-induced prompt $\gamma$-radiation.

The absolute efficiency in the counting geometry was determined using the $^{12}$C(p,$\gamma$)$^{13}$N reaction, as described in \cite{Gyurky2019}. After the determination of the efficiency in a reference geometry with calibration sources, a $^{12}$C target was irradiated. The roughly 10 min half-life of $^{13}$N gives the opportunity to measure its decay in the counting geometry (the target is inside the target chamber) and then in the reference geometry (the target is fixed on the detector end cap). From the comparison of the two measurements, the absolute efficiency in the counting geometry can be calculated, which was found to be $\eta$\,=\,1.83\,$\pm$\,0.05\%.

\section{\label{sec:analysis} Data analysis and cross section results}

\begin{figure}[ht]
   \includegraphics[clip,width=1\linewidth]{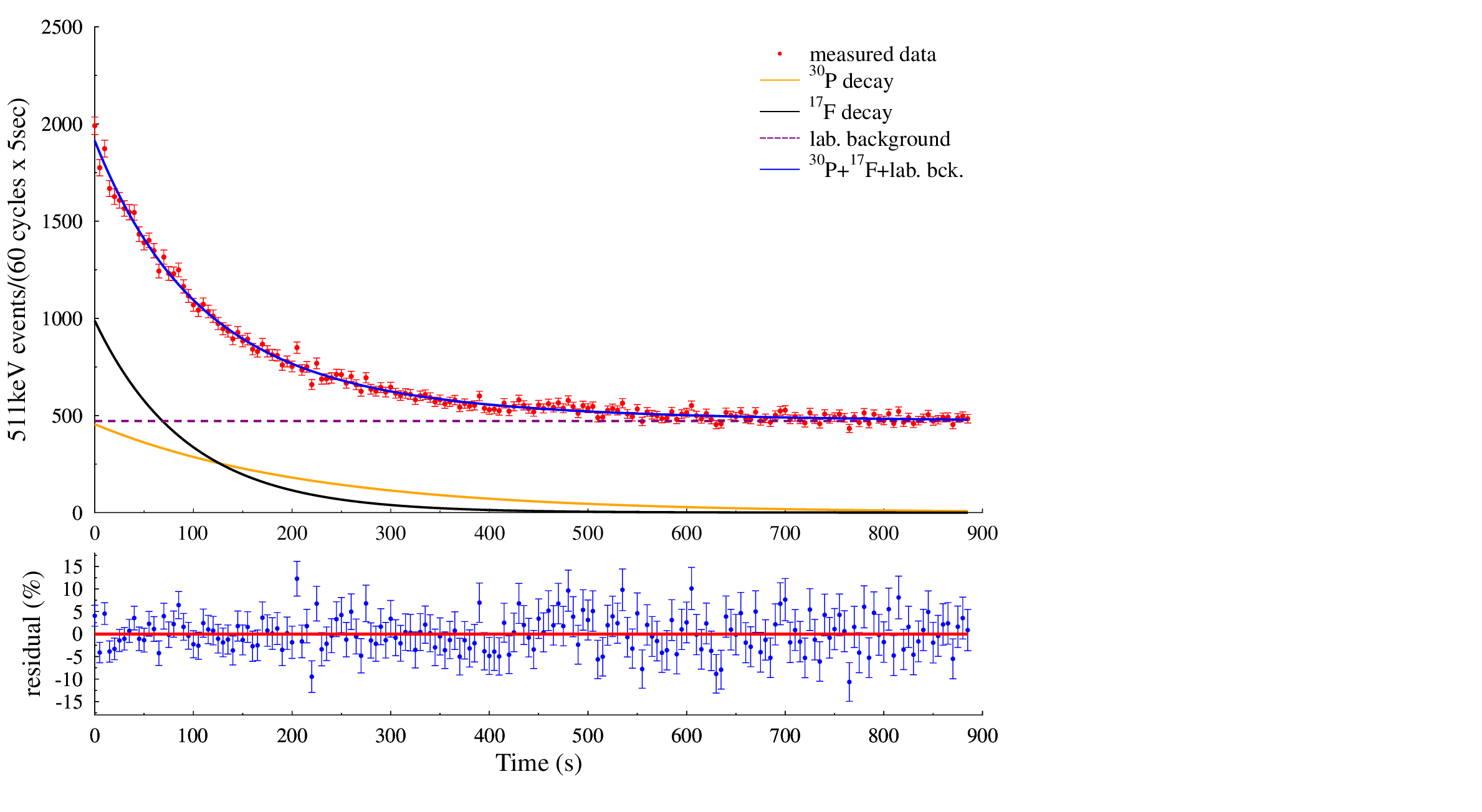}
    \caption{Number of detected 511\,keV $\gamma$-rays as a function of time and the fit of the data including the three components as described in the text. The lower panel shows the percentage fit residuals.}
    \label{fig:decay}
\end{figure}

The 20 min cycles of a run were added up in order to facilitate the fit of the decay curve and thus to obtain the cross section. An example can be seen in fig \ref{fig:decay} which shows the sum of 60 cycles of run \#10. The 511\,keV radiation does not only originate from the decay of the $^{30}$P reaction product, but also from the laboratory $\gamma$-background and from other nuclear reactions producing radioactive isotopes, which decay by positron emission. Since the target layers contain significant amount of oxygen, such a disturbing reaction is $^{16}$O(p,$\gamma$)$^{17}$F. The half-life of $^{17}$F is 64.49\,$\pm$\,0.16 sec \cite{Tilley1993}. The different half-lives of $^{30}$P and $^{17}$F allows the separation of the two isotopes by fitting two exponential functions to the measured decay curve. For such a fit, the decay constants of both isotopes can be kept fixed, as the half-lives are precisely known in the literature. Thus, only a time-independent laboratory background and the number of decaying $^{30}$P and $^{17}$F isotopes are the fit parameters, allowing the determination of the number of reactions and thus the cross sections. The three components of the fit are also plotted separately in fig \ref{fig:decay}, as well as the fit residuals in the lower panel. 

Knowing the areal density of the $^{16}$O and $^{29}$Si atoms in the target, the cross section of both $^{16}$O(p,$\gamma$)$^{17}$F and $^{29}$Si(p,$\gamma$)$^{30}$P reactions could be determined from the fits discussed above. The obtained cross sections are listed in table \ref{tab:results} for all 13 runs. The first column shows the effective proton energies which were calculated by taking into account the thickness of the targets. In the investigated 1000\,keV-1430\,keV energy range, the proton beam loses typically 11-15\,keV in the targets (based on SRIM calculations \cite{SRIM}). As we have no a priori information about the energy dependence of the cross section within such energy intervals, the effective energies were assigned to the middle of the targets (half energy loss) and the quoted uncertainties represent the whole energy range covered by the targets. The quoted cross section uncertainties are statistical only, obtained from the fit of the decay curves. The last column shows the reduced chi square values of the fits. The number of degrees of freedom in each fit was 175 corresponding to 3 fit parameters and 178 measured data points (5 sec intervals during the 15 min decay counting, excluding the first and last intervals). The $\chi^2_{\rm red.}$ values are always close to one, indicating the good fit obtained with the sum of two exponentials and a background. The absence of any structure in the fit residuals (lower panel in fig \ref{fig:decay}) also proves the good fit. 

The runs carried out at a given energy can be combined to obtain the final cross section. These final results are summarized in table \ref{tab:results2} for both $^{29}$Si(p,$\gamma$)$^{30}$P and $^{16}$O(p,$\gamma$)$^{17}$F reactions, where the astrophysical S-factors\footnote{The astrophysical S-factor is related to the reaction cross section but removes its strong energy dependence due to the Coulomb-barrier penetration. For its definition see \cite{Iliadis2015}.} calculated from the cross sections are also listed. In this table, the cross section (and S-factor) uncertainties are the total ones, including the statistical errors (after averaging the runs at the same energies) and the systematic uncertainties. The latter include the target thickness (7.5\,\%), detection efficiency (3\,\%) and current integration (3\,\%) uncertainties. 

\begin{table}
\caption{\label{tab:results} Measured cross sections of the $^{29}$Si(p,$\gamma$)$^{30}$P and $^{16}$O(p,$\gamma$)$^{17}$F reactions. The quoted cross section uncertainties are statistical only.} 
\begin{indented}
\item[]\begin{tabular}{lr@{\hspace{1mm}}c@{\hspace{1mm}}lr@{\hspace{1mm}}c@{\hspace{1mm}}lr@{\hspace{1mm}}c@{\hspace{1mm}}lc}
\br
Run & \multicolumn{3}{c}{$E_{\rm p}^{\rm eff}$ (keV)} & 		\multicolumn{6}{c}{cross section ($\mu$barn)}	& $\chi^2_{\rm red.}$	 \\
& & & 				&			\multicolumn{3}{c}{$^{29}$Si(p,$\gamma$)$^{30}$P} & \multicolumn{3}{c}{$^{16}$O(p,$\gamma$)$^{17}$F} \\
\mr
\#1	&	1092.7	&	$\pm$	&	7.2	&	0.68	&	$\pm$	&	0.26	&	1.58	&	$\pm$	&	0.31	&	0.89	\\
\#2	&	992.4	&	$\pm$	&	7.5	&	0.30	&	$\pm$	&	0.11	&	1.32	&	$\pm$	&	0.09	&	1.08	\\
\#3	&	1240.3	&	$\pm$	&	6.7	&	1.73	&	$\pm$	&	0.18	&	2.55	&	$\pm$	&	0.16	&	1.11	\\
\#4	&	1421.3	&	$\pm$	&	6.1	&	2.55	&	$\pm$	&	0.19	&	2.98	&	$\pm$	&	0.42	&	0.93	\\
\#5	&	1421.7	&	$\pm$	&	5.7	&	1.86	&	$\pm$	&	0.20	&	3.50	&	$\pm$	&	0.12	&	0.84	\\
\#6	&	992.9	&	$\pm$	&	7.0	&	0.85	&	$\pm$	&	0.26	&	1.06	&	$\pm$	&	0.13	&	0.94	\\
\#7	&	1093.2	&	$\pm$	&	6.7	&	0.99	&	$\pm$	&	0.16	&	1.59	&	$\pm$	&	0.13	&	0.79	\\
\#8	&	1241.0	&	$\pm$	&	6.0	&	1.43	&	$\pm$	&	0.14	&	2.62	&	$\pm$	&	0.15	&	1.16	\\
\#9	&	1093.4	&	$\pm$	&	6.5	&	0.36	&	$\pm$	&	0.20	&	2.15	&	$\pm$	&	0.22	&	1.15	\\
\#10	&	1423.9	&	$\pm$	&	5.5	&	2.08	&	$\pm$	&	0.10	&	3.66	&	$\pm$	&	0.12	&	0.94	\\
\#11	&	1093.4	&	$\pm$	&	6.5	&	0.82	&	$\pm$	&	0.16	&	1.89	&	$\pm$	&	0.17	&	1.00	\\
\#12	&	1241.0	&	$\pm$	&	6.0	&	1.46	&	$\pm$	&	0.14	&	3.08	&	$\pm$	&	0.16	&	1.12	\\
\#13	&	1423.9	&	$\pm$	&	5.5	&	1.89	&	$\pm$	&	0.24	&	4.18	&	$\pm$	&	0.26	&	0.87	\\
\br
\end{tabular}
\end{indented}
\end{table}

\begin{table}
\caption{\label{tab:results2} Final cross section results of $^{29}$Si(p,$\gamma$)$^{30}$P and $^{16}$O(p,$\gamma$)$^{17}$F. The quoted total uncertainties include the statistical and systematic ones. The astrophysical S-factors calculated from the cross sections are also listed.} 
\begin{indented}
\item[]\begin{tabular}{lr@{\hspace{1mm}}c@{\hspace{1mm}}lr@{\hspace{1mm}}c@{\hspace{1mm}}lr@{\hspace{1mm}}c@{\hspace{1mm}}l}
\br
Runs & \multicolumn{3}{c}{$E_{\rm p}^{\rm eff}$ (keV)} & \multicolumn{3}{c}{cross section ($\mu$barn)} &	\multicolumn{3}{c}{S-factor (keV$\cdot$barn)} \\
\mr
\multicolumn{10}{c}{$^{29}$Si(p,$\gamma$)$^{30}$P}\\
\#2, \#6	&	992.6	&	$\pm$	&	7.5	&	0.38	&	$\pm$	&	0.20	&	421	&	$\pm$	&	218	\\
\#1, \#7, \#9, \#11	&	1093.2	&	$\pm$	&	7.2	&	0.77	&	$\pm$	&	0.15	&	486	&	$\pm$	&	95	\\
\#3, \#8, \#12	&	1240.8	&	$\pm$	&	6.7	&	1.51	&	$\pm$	&	0.16	&	482	&	$\pm$	&	50	\\
\#4, \#5, \#10, \#13	&	1422.7	&	$\pm$	&	6.1	&	2.11	&	$\pm$	&	0.22	&	338	&	$\pm$	&	36	\\
\mr
\multicolumn{10}{c}{$^{16}$O(p,$\gamma$)$^{17}$F}\\
\#2, \#6	&	992.6	&	$\pm$	&	7.5	&	1.23	&	$\pm$	&	0.16	&	3.37	&	$\pm$	&	0.44	\\
\#1, \#7, \#9, \#11	&	1093.2	&	$\pm$	&	7.2	&	1.77	&	$\pm$	&	0.20	&	3.64	&	$\pm$	&	0.40	\\
\#3, \#8, \#12	&	1240.8	&	$\pm$	&	6.7	&	2.74	&	$\pm$	&	0.29	&	4.03	&	$\pm$	&	0.42	\\
\#4, \#5, \#10, \#13	&	1422.7	&	$\pm$	&	6.1	&	3.62	&	$\pm$	&	0.34	&	3.80	&	$\pm$	&	0.35	\\
\br
\end{tabular}
\end{indented}
\end{table}

\section{\label{sec:discussion} Discussion and conclusions}

\begin{figure}[ht]
   \includegraphics[clip,width=1\linewidth]{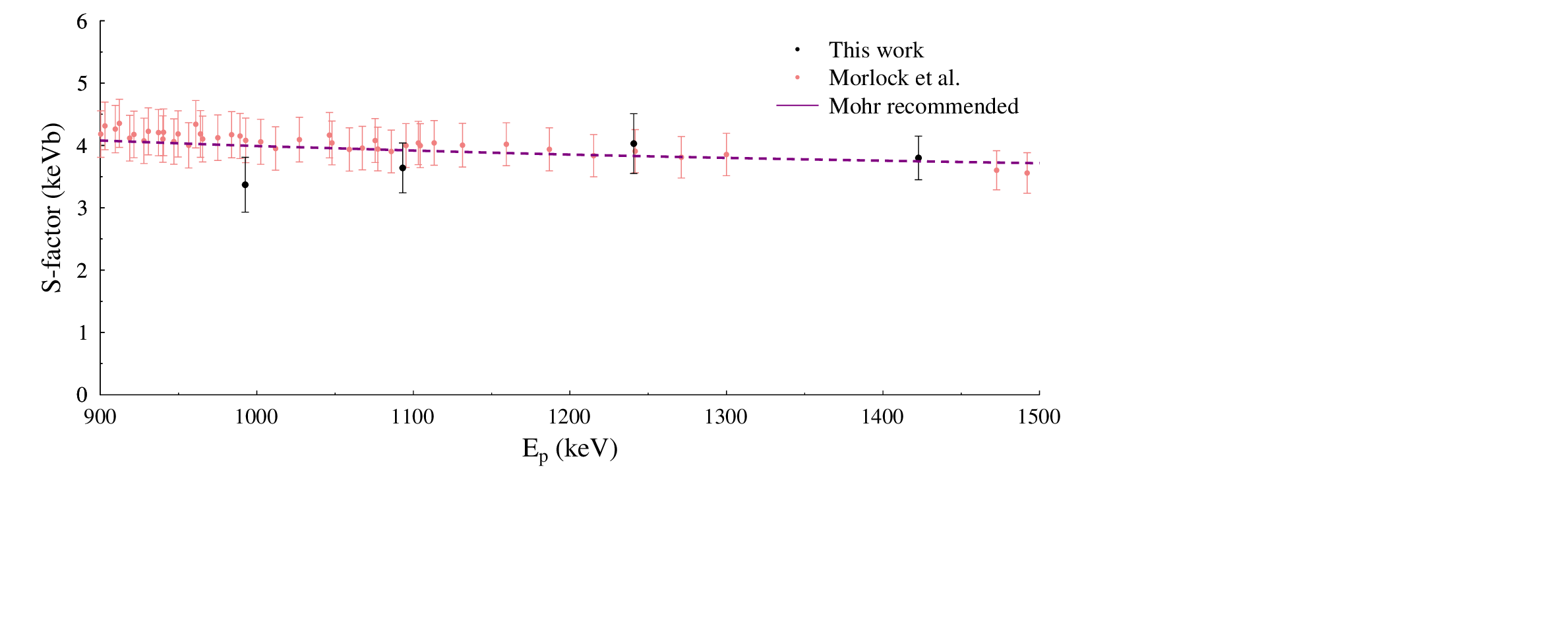}
    \caption{Measured S-factor of the $^{16}$O(p,$\gamma$)$^{17}$F reaction compared with the data of Morlock et al. \cite{Morlock1997} and with the recommended S-factors of Mohr and Iliadis \cite{Mohr2012}. The plotted error bars correspond to the total uncertainty.}
    \label{fig:O_result}
\end{figure}

\begin{figure}[ht]
   \includegraphics[clip,width=1\linewidth]{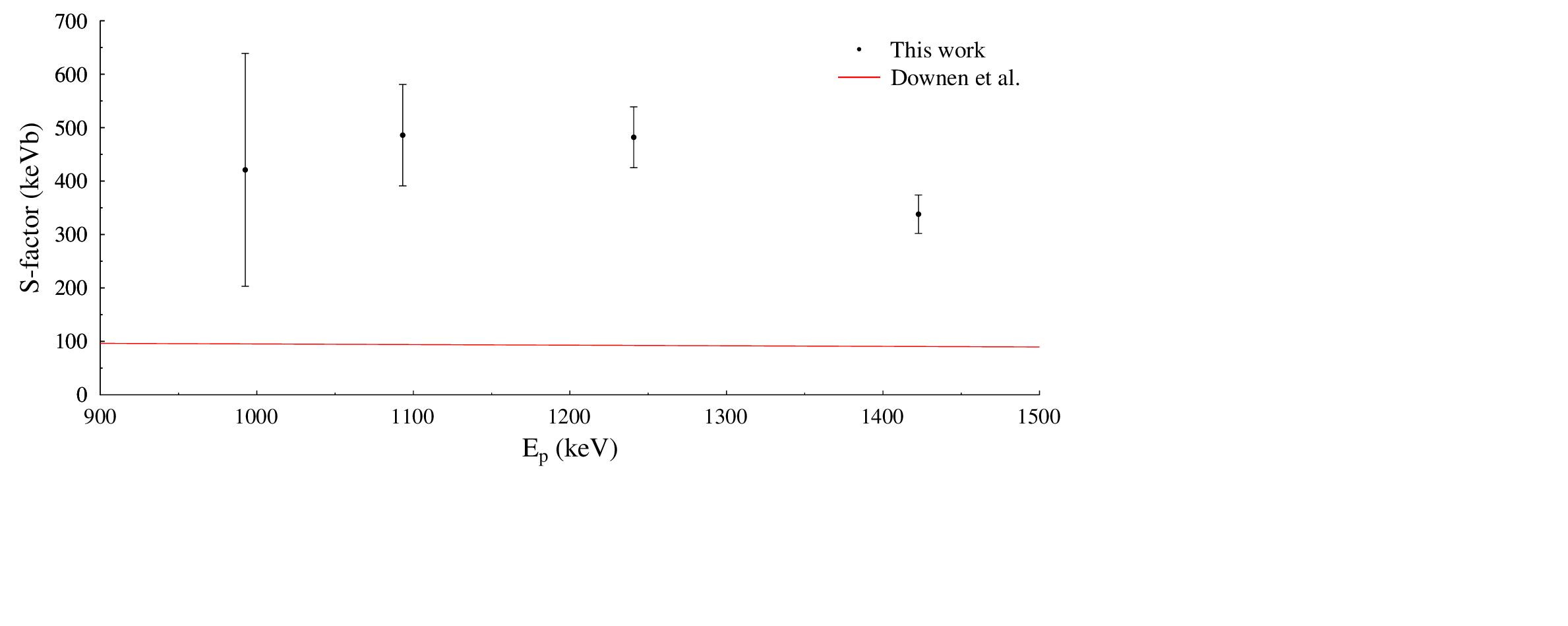}
    \caption{Measured S-factor of the $^{29}$Si(p,$\gamma$)$^{30}$P reaction compared with the theoretical S-factor curve of Downen et al. \cite{Downen2022}. The plotted error bars correspond to the total uncertainty.}
    \label{fig:Si_result}
\end{figure}

The S-factors listed in table \ref{tab:results2} are also plotted in figs \ref{fig:O_result} (for $^{16}$O(p,$\gamma$)$^{17}$F) and \ref{fig:Si_result} (for $^{29}$Si(p,$\gamma$)$^{30}$P). The $^{16}$O(p,$\gamma$)$^{17}$F cross section is relatively well known in the literature in our studied energy range. Fig \ref{fig:O_result} shows the total cross section (i.e. the sum of the two partial cross sections leading to the ground state and first excited states in $^{17}$F) measured by Morlock et al. \cite{Morlock1997}. The figure also shows the recommended cross section by Mohr and Iliadis \cite{Mohr2012} as a solid line. Owing to the larger uncertainty, our data are less precise than the literature values. However, the good agreement of our results with the literature data provides an indirect proof that the target characterizations, the double exponential fit of the decay curves and thus the obtained cross sections are reliable.

As it is emphasized in section \ref{sec:intro}, for the $^{29}$Si(p,$\gamma$)$^{30}$P reaction there was no DC cross section available in the literature before the present work. Therefore, in fig \ref{fig:Si_result} the present results are compared only with a theoretical S-factor curve. This curve is based on the work of Downen et al. who used a potential model and the measured proton spectroscopic factor for calculating the DC cross section \cite{Downen2022}. The energy dependence of our S-factor values are consistent with that of the theoretical curve (almost constant S-factor in this energy range) but our S-factors are significantly higher than the calculated ones. Adopting the slope of the theoretical S-factor function, the error-weighted average deviation of our data from the calculation is a factor of 4.3\,$\pm$\,0.6. It is worth noting that -- owing to the experimental constraints discussed above -- our measurement was carried out in a relatively narrow energy range. Therefore, theoretical considerations cannot be avoided when the nonresonant cross section is needed at lower energies. Nevertheless, our data provides an experimental constraint to theoretical models at the studied energies. 

As it is already stated above, it cannot be excluded that besides the direct capture, our measured cross section includes contribution from resonances. High-energy broad resonances may have non-negligible cross section in our studied energy range. Weak narrow resonances - not observed in previous experiments - may also be present near the energies of our measurements. However, the observed fairly constant S-factor indicate that mainly the DC cross section is measured. The cross section values much higher than the estimate of Downen et al. call for further experimental data as well as new theoretical calculations. This latter is beyond the scope of the present experimental work. $^{29}$Si(p,$\gamma$)$^{30}$P is not the only reaction relevant to nova nucleosynthesis where DC cross section is solely based on theoretical calculations. In the present work, significant deviation was found between theory and experiment. This may have an impact on the Si isotopic ratios. If other reactions were to exhibit similar deviations, it would influence strongly the nucleosynthesis model calculations \cite{IliadisJose_priv}. DC cross section measurements of further reactions in this mass range is thus recommended.

\ack
The authors thank C. Iliadis and J. Jos\'e for useful discussions. This work was supported by NKFIH grant K134197, by the \'UNKP-23-3 New National Excellence Programs of the Ministry for Culture and Innovation from the source of the National Research, Development and Innovation fund and by the Hungarian Government, Economic Development and Innovation Operational Programme (GINOP-2.3.315-2016-00005) grant, co-funded by the EU.

\section*{References}

\providecommand{\newblock}{}

\end{document}